\documentclass[twocolumn]{article}
\usepackage[a4paper, margin=1.8cm]{geometry}

\usepackage{graphicx} 
\usepackage{lipsum}
\usepackage{hyperref}
\usepackage{algorithm,algpseudocode}
\usepackage{xcolor}
\usepackage{amsmath, amssymb, amsfonts, amsthm}
\usepackage{siunitx}
\usepackage{balance}

\newtheorem{remark}{Remark}
\newtheorem{result}{Result}
\newtheorem{theorem}{Theorem}

\newcommand{\R}{\mathbb{R}}
\newcommand{\prox}{{\rm prox}}
\newcommand{\norm}[1]{\left\Vert #1 \right\Vert}
\newcommand{\C}{\mathcal{C}}

\title{Real-Time Single-Iteration Model Predictive Control for Wave Energy Converters}
\author{S. Pirrera, \and N. Faedo, \and S. M. Fosson, \and  D. Regruto
\thanks{Nicolas Faedo is within the Department of Mechanical and Aerospace Engineering, Politecnico di Torino. Sophie M. Fosson, Simone Pirrera and Diego Regruto are with the Department of Control and Computer Engineering, Politecnico di Torino; 
e-mail: \{nicolas.faedo, sophie.fosson, simone.pirrera, diego.regruto\}@polito.it. Corresponding author: \textit{S. Pirrera}.
}}

\date{}

\begin{document}

\maketitle

\begin{abstract}
This paper proposes a novel real-time algorithm for controlling wave energy converters (WECs). We begin with the economic model predictive control (MPC) problem formulation and apply a novel, first-order optimization algorithm inspired by recently developed control-based algorithms for constrained optimization to define the controller dynamics according to the single-iteration MPC approach. We theoretically analyse the convergence of the employed algorithm and the computational complexity of the obtained controller. Results from simulations using a benchmark WEC system indicate that the proposed approach significantly outperforms standard MPC, thanks to the inherent ability to handle faster sampling rates.
\end{abstract}

\section{Introduction}

Ocean energy offers a renewable energy source with outstanding energy generation potential, able to provide a stable and predictable baseload, with an absence of night-day variability, and seasonal complementarity with solar and wind \cite{bhattacharya2021timing, kluger2023power}. Within ocean energy, wave energy is essential to diversify the energy mix and ensure a balanced energy supply, having the massive potential of 3,2 MWh/year \cite{mccormick2013ocean}. Nonetheless, wave energy remains virtually unexploited due to the high levelized cost of energy (LCoE), which currently discourages large investments in this technology \cite{guo2021review}. To lower the levelized cost of energy, control technology plays a crucial role \cite{ringwood2023empowering}, as improved control actions can significantly enhance the energy generation capabilities of wave energy converters (WECs).

The current mainstream approach to optimal WEC control relies on economic model predictive control (MPC) \cite{faedo2017optimal, ringwood2023empowering}, a paradigm according to which the control input is computed as the solution to an optimization problem seeking to optimize a given objective (\emph{i.e.} maximize energy absorption from the incoming wave field), subject to input and state constraints representing device physical limtiations, at each sampling rate, in a receding horizon fashion. This choice is strongly motivated by the flexibility of the approach, its intrinsic capacity to circumvent the standard non-causality that characterizes the WEC control problem, as well as its strong theoretical guarantees~\cite{angeli12}.
%

MPC has been considered and applied to WECs for over a decade, with \cite{gieske} being the first published instance, applying linear MPC to the Archimedes wave swing \cite{de2006modelling}. Since then, several variants of MPC have been formulated for the WEC case, including, \emph{e.g.}, \cite{brekken2011model,li2014model,kody2019model,liao2021high,zhan2025terminal}, which contemplate diverse conversion concepts, model assumptions, and formulation of state and input constraints. Nonetheless, independent of the specific formulation, MPC performances are naturally limited by the need to discretize the WEC model to propagate (\emph{i.e.} predict with the receding window) the device dynamics. This process introduces an approximation (discretization) error, which affects the quality of the prediction and, consequently, can significantly impact energy-generation performance and closed-loop properties. The most straightforward approach to reducing this error is the adoption of higher sampling rates, which comes at the cost of using more expensive computational resources. If these resources are not available, real-time implementation of the controller is precluded, directly compromising its overall role in minimizing the associated LCoE. 

To effectively reduce the sampling rate and avoid the requirement of expensive computational resources, it is necessary to accelerate computations by adopting efficient optimization algorithms. Several algorithms have been proposed to accelerate the solution of quadratic programs (QP) arising in the context of MPC; see, \emph{e.g.}, \cite{her19,stellato_osqp_2020,cal24} for recent contributions in this direction. In this study, we focus on algorithms arising from a recent paradigm based on a control-design approach, named controlled multipliers optimization~\cite{cerone_new_2025}. Specifically, we start from the feedback-linearization CMO (FL-CMO) algorithm, which enjoys a faster convergence rate compared to standard alternatives such as the primal-dual gradient dynamics~\cite{qu_exponential_2019}, and extend it to define a novel optimization algorithm named Proj-FL-CMO. Unlike the recent extensions of CMO in \cite{centorrino_proximal_2025} and \cite{cerone_feedback_2024}, Proj-FL-CMO is defined in discrete time, ensuring a direct physical interpretation in terms of execution speed on digital devices. Moreover, we theoretically analyze it to prove that it converges globally for certain non-convex problems.

Next, motivated by the fundamental role that efficient MPC
formulations can play in the pathway towards commercialization of WEC systems, and inspired by recent results on \textit{single-iteration MPC} \cite{emiliano25}, we define a novel dynamic controller for WECs by adopting the developed Proj-FL-CMO algorithm. The single-iteration MPC approach amounts to replacing the MPC controller (which requires the solution of a full optimization problem) with a single iteration of the algorithm used to solve the same optimization problem. This approach is strongly related to the online feedback optimization paradigm, where the controller is defined based on dynamics that converge to the solution of a suitable optimization problem; see, \emph{e.g.},~\cite{col20,haus21}. Also, the approach is connected with the {real-time iterations} (RTI) paradigm, well known in the context of nonlinear MPC, with the key difference that RTI amounts to the solution, at each time step, of a full QP representing the single iteration of a sequential QP algorithm for solving the nonlinear MPC problem, see \emph{e.g.} \cite{DIEHL2002,gros2020}. Within this study, we analyze the computational complexity of the defined dynamic controller and establish the maximum achievable sampling frequency. Finally, we validate the performance of the overall controller on a benchmark WEC system, demonstrating that the approach significantly enhances energy production compared to standard MPC using the same computational resources.

The remainder of this paper is organized as follows. Section \ref{sec:problem_formulation} formulates the WEC control problem using a standard MPC algorithm, and presents an analysis of the computational resources required for its execution. Section \ref{sec:projflcmo} presents the core of our study, \emph{i.e.} the novel Proj-FL-CMO algorithm, providing guarantees of convergence and insight on parameter selection. Section \ref{sec:mpc_design} presents the re-formulation of the standard MPC using the proposed novel optimization technique, leading to a single-iteration MPC scheme suitable for real-time control of WEC systems. Finally, Section \ref{sec:simulation} offers a numerical appraisal of the overall proposed single-iteration MPC, applied to a benchmark WEC system, demonstrating its performance in terms of energy absorption, while Section \ref{sec:conclusions} encompasses the main conclusions of our study.  

\section{Problem Formulation}
\label{sec:problem_formulation}

This section introduces the standard formulation of MPC for single-degree-of-freedom (DoF) WEC systems\footnote{The case of multi-DoF can be treated analogously, see \emph{e.g.} \cite{li2014model2}.}. Consistently, we model the WEC device following linear potential flow theory assumptions, \emph{i.e.} in terms of the continuous-time linear time-invariant dynamical system 
\begin{equation}
\label{eq:ss}
\mathcal{P}:\left\{ \begin{aligned}
      \dot x &= A_c x + B_c (u+w),\\
      y &= C_c x,
\end{aligned}\right.
\end{equation}
where $u(t)\in\mathbb{R}$ is the control input force, $w(t)\in\mathbb{R}$ is the wave excitation force generated by the surrounding wave field, and $y(t) = [p(t)\,\, v(t)]^{\top} \in\mathbb{R}^2$ is a vector containing position and velocity (assumed measurable, consistent with standard WEC experimental setups). The triple $(A_c, B_c, C_c) \in \mathbb{R}^{n\times n} \times \mathbb{R}^{n} \times \mathbb{R}^{2\times n}$ can be either computed directly from the so-called Cummins' equation \cite{cummins1962impulse}, leveraging boundary element methods \cite{papillon2020boundary}, or via data-based (system identification) procedures \cite{pasta2023towards} (see also the discussion in Section \ref{sec:simulation}).
\begin{remark}
\label{rem:system_properties}
    Due to the physical properties of the wave energy conversion process, system $\mathcal{P}$ in \eqref{eq:ss} is internally stable, minimum-phase, and passive (see \emph{e.g.} \cite{scruggs2010causal}).
\end{remark}

To apply the standard MPC setup, we describe system \eqref{eq:ss} in discrete-time through a zero-order hold (ZOH) discretization with step-size $T$, obtaining
\begin{equation}
\label{eq:ss_d}
\begin{aligned}
    x^+ &= A x + B (u+w), \\
    p &= C_p x, ~v = C_v x 
\end{aligned}
\end{equation}
\begin{remark}\label{rmk:bad_discr}
    System \eqref{eq:ss_d} does not perfectly describe the WEC plant \eqref{eq:ss}, since the ZOH assumption holds true only for the control input $u$, while $w$ changes continuously, not in discrete steps with period $T$.
\end{remark}  

Leveraging \eqref{eq:ss_d} we define the standard MPC problem for WECs (see \emph{e.g.} \cite{li2014model}) as
\begin{subequations}\label{eq:mpc_original}\begin{align}
    & \min_{\hat u,\hat p,\hat v \in \R^{N}, \hat x \in \R^{nN}}  ~~\sum_{i=1}^N \hat u_i \hat v_i + \frac{1}{2} r \hat u^2_i\\
    &\textrm{s.t.}  \quad \hat x_1 = x_k \\
    &\hat x_{i+1} = A \hat x_i + B( \hat u_i + w_i), ~~ i = 1,\dots, N-1 \\
    &\hat p_{i} = C_p \hat x_i, ~~\underline{p} \leq \hat p_i \leq \overline{p},  ~~ i = 1,\dots, N \\
    &\hat v_{i} = C_v \hat x_i, ~~ \underline{v} \leq \hat v_i \leq \overline{v}, ~~ i = 1,\dots, N \\
    &\underline{u} \leq \hat u_i \leq \overline{u}, ~~ i = 1,\dots, N.
\end{align} \end{subequations}
We can recursively eliminate variables $\hat x, \hat v, \hat p$, through the relations
\begin{subequations}\label{relations}\begin{align}
    \hat x &= \mathcal{A} x_k + \mathcal{B} (\hat u + W_k), \\
    \hat p &= \mathcal{C}_{x,p} x_k + \mathcal{C}_{u,p} (\hat u + W_k),\\
    \hat y &= \mathcal{C}_{x,v} x_k + \mathcal{C}_{u,v} (\hat u + W_k),
\end{align}\end{subequations}
where \begin{align}
    &\mathcal{A} = \begin{bmatrix} I \\ A \\ A^2 \\ \vdots \\ A^{N-1} \end{bmatrix}, ~~ \mathcal{B} = \begin{bmatrix} 0 & 0 & \dots & 0 \\ B & 0 & \dots & 0 \\ AB & B & \dots & 0 \\ \vdots & \vdots &\vdots &\vdots \\ A^{N-2}B & A^{N-3}B & \dots & 0 \end{bmatrix},\\
    &\mathcal{C}_{x,p} = (I \otimes C_{p})\mathcal{A}, ~~ \mathcal{C}_{u,p} = (I \otimes C_{p})\mathcal{B},\\
    &\mathcal{C}_{x,v} = (I \otimes C_{v})\mathcal{A}, ~~ \mathcal{C}_{u,v} = (I \otimes C_{v})\mathcal{B},
\end{align}
$\otimes$ is the standard Kronecker product, $I$ denotes the identity matrix (dimensioned according to the context), and $W_k \doteq [w_k, \dots, w_{k+N-1}]^\top \in \R^{N}$ is the current wave prediction\footnote{Note that we use the term "wave prediction" to denote both instantaneous and future values of $w$ within a given receding window.} along the horizon $N$. These relationships put the MPC problem in the standard quadratic programming (QP) form:
\begin{equation}\begin{aligned}
    \min_{\hat u \in \R^N}& ~~\hat u^\top \left(\frac{r}{2}I + \mathcal{C}_{u,v}\right) \hat u + \left(\mathcal{C}_{x,v} x_k + \mathcal{C}_{u,v} W_k \right)^\top \hat u \\
    & \textrm{s.t.}~~\begin{bmatrix}
        \phantom{-}\mathcal{C}_{u,p} \\ -\mathcal{C}_{u,p}\\ \phantom{-}\mathcal{C}_{u,v} \\ -\mathcal{C}_{u,v}
    \end{bmatrix} \hat u \leq \begin{bmatrix}
        \phantom{-}\overline{p} - \mathcal{C}_{u,p} W_k - \mathcal{C}_{x,p} x_k \\ -\underline{p} + \mathcal{C}_{u,p} W_k + \mathcal{C}_{x,p} x_k \\ \phantom{-}\overline{v} - \mathcal{C}_{u,v} W_k - \mathcal{C}_{x,v} x_k \\ -\underline{v} + \mathcal{C}_{u,v} W_k + \mathcal{C}_{x,v} x_k
    \end{bmatrix}.
\end{aligned}\end{equation}
As shown in~\cite{li_model_2014,papini2024frequency}, if $r>0$ is sufficiently large, the problem is convex, thus admits a unique global optimum. 

\begin{remark}\label{rmk:need_small_T}
    The impact of the discretization step size $T$ on the overall WEC controller performance is twofold. Firstly, since the original continuous-time system is passive (see Remark \ref{rem:system_properties}), a smaller $r$ is required when decreasing $T$, thus making the contribution of the energy in the MPC cost more relevant. Secondly, following Remark~\ref{rmk:bad_discr} and assuming availability of perfect wave information, the approximation error induced by the (incorrect) ZOH discretization for the input $w$ reduces, making the prediction more precise and improving the overall performance.
\end{remark}

\subsection{Computational Resources Analysis}

Following Remark~\ref{rmk:need_small_T}, we aim to get an MPC-like algorithm compatible with the selection of a small $T$. However, this is associated with a rapidly growing computational effort, which hinders the approach's real-time capabilities. In the following, we make the following standard choice in experimental scenarios (see \emph{e.g.} \cite{ringwood2023wave}): the prediction horizon $N$ of the MPC is according to 
\begin{equation}
    N = \frac{T_p}{T},
\end{equation}
where $T_p$ is the continuous-time horizon with which it is possible to confidently predict the incoming wave force $w$. Since $T_p$ is a fixed time interval, $N$ grows linearly with the increase in the sampling frequency. In turn, the size of the QP increases, as it consists of $N$ optimization variables and $4N$ inequality constraints. Using a standard interior-point method, the required amount of computations (measured in floating point operations, FLOPs) can be estimated as $c_{tot} = n_i c_i$, where $n_i = O(\sqrt{\kappa} \log(1/\epsilon))$ is the number of required iterations to achieve accuracy $\epsilon$, $\kappa$ denotes the condition number of the KKT system matrix, and $c_i = O(5 N^3)$ is the cost of each iteration after performing the KKT system reduction via Shur complement; see \cite{nocedal2006numerical} for further details.

Let OP denote the hardware's capability, measured in FLOP/s. Overall, a reliable estimate for QP computation delay $\Delta_\mathrm{ipm}$ satisfies:
\begin{equation}
    \Delta_\mathrm{ipm} = \frac{n_i 5 N^3}{OP}, \quad \text{with} \quad N = \frac{T_p}{T}.
\end{equation}
Since we must ensure $\Delta_\mathrm{ipm} \ll T$, a reasonable estimate for the minimum sampling rate is obtained by enforcing $T = 10 \Delta_\mathrm{ipm}$, which leads to:
\begin{equation}\label{eq:estim_T_ipm}
    T = \sqrt[4]{\frac{50 n_i T_p^3}{\rm OP}}
\end{equation}
\emph{i.e.}, reducing the sampling rate by a factor $\alpha$ is related to an increase in the demand for computational resources by a factor of $\alpha^4$. 

\section{Proj-FL-CMO Algorithm}\label{sec:projflcmo}

This section introduces and analyzes the algorithm used in the single-iteration MPC design, presented and developed within Section~\ref{sec:mpc_design}. Let us consider an optimization problem of the form
\begin{subequations}\label{eq:opt_probl}
\begin{align}
    \xi^\star &= \arg\min_\xi f(\xi) \\
    & \quad \text{s.t.}\quad h(\xi)=0, \\
    & \qquad \underline{b} \leq \xi \leq \overline{b} \label{eq:box_cns_generic}.
\end{align}\end{subequations}
We define the \textit{projected feedback linearization controlled multipliers optimization} (Proj-FL-CMO) algorithm as:
\begin{subequations}\label{eq:projFLCMO}\begin{align}
    \xi^+ &= \Pi(\xi - \tau \nabla f(\xi) - \tau J_h^\top \lambda), \label{eq:xi_update} \\
    \lambda &= (J_hJ_h^\top)^{-1}(-J_h\nabla f(\xi) - k_p h(\xi) - k_i z), \\
    z^+ &= z + \tau h(\xi),
\end{align}\end{subequations}
where $\Pi(\cdot)$ denotes the projection operator in the box defined by constraints \eqref{eq:box_cns_generic}, \emph{i.e.} $\Pi(\xi) \doteq \min(\overline{b}, \max(\underline{b}, \xi))$.

The discrete-time dynamical system \eqref{eq:projFLCMO} is inspired by the FL-CMO dynamics originally proposed in \cite{cerone_new_2025}, \emph{i.e.}
\begin{subequations}\label{eq:FLCMO}\begin{align}
    \dot x &= -\nabla f(x) - J_h^\top \lambda \\
    \lambda &= (J_hJ_h^\top)^{-1}(-J_h\nabla f(x) +v(t) )
\end{align}\end{subequations}
where $v(t)$ is determined by the proportional-integral (PI) law $v(t) = \mathcal{G}(h(\xi)) = k_p h(\xi) + k_i \int_{0}^t h(\xi(\tau)) {\rm d} \tau$. Specifically, the algorithm \eqref{eq:projFLCMO} is obtained after forward Euler discretization of the differential equation \eqref{eq:FLCMO} with step-size $\tau \in \R_+$, and introducing a discontinuous projection step in the definition of the optimization variable update \eqref{eq:xi_update}. 

Although extensions of the CMO framework enabling the handling of inequality constraints were recently proposed in \cite{cerone_feedback_2024} and \cite{centorrino_proximal_2025}, unlike both algorithms proposed in such studies, Proj-FL-CMO enjoys global exponential convergence despite the non-strong convexity of $f(\xi)$. Moreover, unlike \cite{cerone_feedback_2024}, convergence is guaranteed despite the linear dependence of the inequality constraints defining the bounds \eqref{eq:box_cns_generic}. 

\subsection{Convergence Analysis}
\label{sec:converg_analysis}

We denote $\iota(\xi)$ the indicator function of the set $\{\xi \in \R^{3N}: \underline{b} \leq \xi \leq \overline{\xi}\}$. It is a classical result that $\Pi(a) = \prox_{\tau \iota}(a)$, where $\prox$ is the proximal operator of the function $\tau \iota$ and $\tau \in \R_+$. For a convex function $g: \R^{n} \rightarrow \R$, we denote as $\partial g \subseteq \R^n$ its subgradient. We are now ready to present the following result.

\begin{result}[Equilibria]\label{result:equil}
    A point $(\xi^\star,z^\star)$ is an equilibrium point of the Proj-FL-CMO dynamics \eqref{eq:projFLCMO} if and only if it satisfies the first-order optimality conditions of Problem \eqref{eq:mpc_eq_and_bnd}.
\end{result}
\begin{proof}
    By definition of equilibrium point, $(\xi^\star,z^\star)$ is an equilibrium if and only if $z^\star = z^\star + \tau h(\xi^\star)$, which is equivalent to primal feasibility: $h(\xi^\star)=0$. Next, consider the first equilibrium equation 
    \begin{equation*}
        \xi^\star = \Pi( \xi^\star - \tau \nabla f(\xi^\star)- \tau J_h^\top(\xi^\star) \lambda ).
    \end{equation*}
    Using the equivalence $a = \Pi(b) = \prox_{\tau \iota}(b) \iff \frac{1}{\tau}(b-a) \in \partial \iota$, we get
    \begin{equation*}
        \frac{1}{\tau}\left( -\xi^\star +\xi^\star - \tau \nabla f(\xi^\star) -\tau J_h^\top(\xi^\star) \lambda \right) \in \partial \iota,
    \end{equation*}
    which leads to the first-order condition
    \begin{equation*}
        0 \in \nabla f(\xi^\star) +\partial \iota + J_h^\top(\xi^\star) \lambda.
    \end{equation*}
\end{proof}
We highlight that Result \ref{result:equil} holds independently of the properties of $f(\xi)$ and $h(\xi)$, which may be non-convex in general.

The following result establishes global exponential convergence under more stringent assumptions. It exploits the quadratic structure of $f$ and the linearity of $h$, but requires less than convexity for $f$, which perfectly matches the setup in Section ~\ref{sec:mpc_design}.

\begin{theorem}[Convergence]\label{th:stab_1}
    Assume that problem \eqref{eq:mpc_eq_and_bnd}, after elimination of all equality constraints, is convex (\emph{i.e.} convex in $\hat u$). Take $\{k_p,k_i\}$ such that $\begin{bmatrix}
        -k_p & -k_i \\ \phantom{-}1 & \phantom{-}0
    \end{bmatrix}$ is Hurwitz and $\tau$ sufficiently small.
    Then, the equilibrium point of the Proj-FL-CMO dynamics is unique and globally exponentially stable with convergence rate $\rho < 1$.  
\end{theorem}
\begin{proof}
    Let $\C^\perp$ be the matrix such that its rows are an orthonormal basis for the null space of the rows of $\C$. Also denote $\delta_\xi = \nabla f(\xi) + \C^\top \lambda$, $\gamma = \C \xi$, and $\eta = \C^\perp \xi$. The following equalities hold.
    \begin{equation}\label{eq:update_xi}\begin{aligned}
        \delta_\xi &= H\xi+f+\C^\top \lambda = \\
        &= \begin{bmatrix}
            \C^\perp \\ \C
        \end{bmatrix}^{-1} \begin{bmatrix}
            \C^\perp H C^{\perp,\top} & \C^\perp H \C^\dagger & 0\\
            0 & k_p I &  k_i I 
        \end{bmatrix}\begin{bmatrix}
            \eta \\ \gamma \\ z
        \end{bmatrix} + \begin{bmatrix}
            \C^\perp f \\ k_p d
        \end{bmatrix}
    \end{aligned}\end{equation}
    where $\C^\dagger$ denotes the Moore-Penrose pseudoinverse of $\C$. Moreover, we notice that \begin{equation}
        \begin{bmatrix}
            \eta \\ \gamma \\ z
        \end{bmatrix} = T^{-1} \begin{bmatrix}
            \xi \\ z
        \end{bmatrix}, \qquad T = \begin{bmatrix}
            \begin{bmatrix}
            \C^\perp \\ \C 
        \end{bmatrix}^{-1} & 0\\
        0 & I \end{bmatrix}.
    \end{equation}

    Define the operator $\Pi_{\rm ext}([a^\top,b^\top]^\top) = [\Pi(a)^\top, b^\top]^\top$. Using \eqref{eq:update_xi} and $z^+ = z +\tau \gamma + \tau d$, the overall update \eqref{eq:projFLCMO} can be written as $\begin{bmatrix} \xi^+\\ z^+\end{bmatrix} = F_{\rm \mathcal{A}}(\xi,z)$, with
    \begin{equation}\label{eq:A_proj_grad_form}
        F_{\rm \mathcal{A}}(\xi,z) \doteq \Pi_{\rm ext} \left( [I - \tau T M T^{-1}] \begin{bmatrix}
            \xi \\ z
        \end{bmatrix} + \begin{bmatrix}
            \C^\perp f \\ k_p d \\ \tau d
        \end{bmatrix}\right). 
    \end{equation}

    To conclude the proof, we show that \eqref{eq:A_proj_grad_form} is strongly contracting, \emph{i.e.} there exists $\rho<1$ such that for all $(\xi_1,z_1), (\xi_2,z_2)$
    \begin{equation}\label{eq:contract_cond}
        \norm{F_{\rm \mathcal{A}}(\xi_1,z_1)-F_{\rm \mathcal{A}}(\xi_2,z_2)} \leq \rho \norm{\begin{bmatrix}
            \xi_1-\xi_2 \\ z_1-z_2
        \end{bmatrix}},
    \end{equation}
    holds. Since the operator $\Pi_{\rm ext}$ is non-expansive, we have
    \begin{equation}
        \norm{F_{\rm \mathcal{A}}(\xi_1,z_1)-F_{\rm \mathcal{A}}(\xi_2,z_2)} \leq \norm{I - \tau T M T^{-1}}_2 \norm{\begin{bmatrix}
            \xi_1-\xi_2 \\ z_1-z_2
        \end{bmatrix}}.
    \end{equation}
    Therefore, the contraction rate $\rho$ satisfies
    \begin{equation}
        \rho \leq \norm{I - \tau T M T^{-1}} < 1.
    \end{equation}
    Since $T M T^{-1}$ is positive definite under the considered assumption on convexity, $k_p$ and $k_i$, under the condition 
    \begin{equation}
    \tau < {\norm{T M T^{-1}}}^{-1},
    \end{equation}
    we have $\rho < 1$.
\end{proof}

\subsection{Parameters Design}\label{sec:param_design}
This section discusses how to design the parameters of $k_p,k_i,\tau$ of \eqref{eq:projFLCMO} to enhance the convergence rate $\rho$. Firstly, we take $\tau$ according to $0.99{\norm{T M T^{-1}}}^{-1}$, to ensure stability and minimally impact performances. Next, we pick
\begin{equation}
    k_p = 2 \norm{\C^\perp H \C^{\perp,\top}}, \quad
    k_i = \norm{\C^\perp H \C^{\perp,\top}}^2
\end{equation}
to enforce  
\begin{equation}
    \lambda_i\left(\begin{bmatrix}
        \phantom{-}k_p & k_i \\ -1 & 0
    \end{bmatrix}\right) = \lambda_{\max}(\C^\perp H \C^{\perp,\top}), \quad \forall i,
\end{equation}
\emph{i.e.}, we place the dynamics of the output coincident with the fastest mode of the zero dynamics for the "inner" FL-CMO design (see \cite{cerone_new_2025} for more details on FL-CMO design). Therefore, for all $i$, there exists $j$ such that $\lambda_i(M) = \lambda_j(\C^\perp H \C^{\perp,\top})$.

\section{Proposed Control Algorithm}
\label{sec:mpc_design}
We propose defining the control input according to the single-iteration MPC paradigm, exploiting the Proj-FL-CMO algorithm to define the iteration. The selection of Proj-FL-CMO, a first-order method, is motivated by its low computational complexity, which makes it feasible to execute the single MPC iteration within reduced time. In the remainder of this section, we define the proposed control algorithm and analyse its computational requirements. 

\subsection{MPC Reformulation}
We start by recasting Problem~\eqref{eq:mpc_original} in the following form:
\begin{subequations}\begin{align}
    & \min_{\hat u,\hat p,\hat y \in \R^{N}}  ~~  \hat u^\top \hat v + \frac{1}{2} r \lVert \hat u\rVert_2^2\\
    &\textrm{s.t.}  \notag \\
    &\hat p = \mathcal{C}_{x,p} x_k + \mathcal{C}_{u,p} (\hat u + W_k)\\
    &\hat v = \mathcal{C}_{x,v} x_k + \mathcal{C}_{u,v} (\hat u + W_k)\\
    &\underline{p} 1_N \leq \hat p \leq \overline{p} 1_N,~\underline{v}1_N \leq \hat v \leq \overline{v}1_N,\\
    &\underline{u} 1_N \leq \hat u \leq \overline{u}1_N.
\end{align} \end{subequations}
Let $\xi = [\hat u,\hat p,\hat v]$. The problem is compactly represented by
\begin{subequations}\label{eq:mpc_eq_and_bnd}\begin{align}
    \min_{\xi \in \R^{3N}}  ~ &  \frac{1}{2} \xi^\top H \xi + f^\top \xi \\
    \textrm{s.t.} \quad &\mathcal{C} \xi + d = 0,\\
    &\underline{b} \leq \xi \leq \overline{b}, \label{eq:bounds_xi}
\end{align} \end{subequations}
where $H,f,\mathcal{C}$ are constant, while 
\begin{equation}\label{eq:d_comput}
    d = \begin{bmatrix} \C_{x,p} \\ \C_{x,v} \end{bmatrix} x_k + \begin{bmatrix} \C_{u,p} \\ \C_{u,v} \end{bmatrix} W_k.
\end{equation}

Note that Problem \eqref{eq:mpc_eq_and_bnd} is not in standard form because we retain variables $\hat p,\hat y$ and equality constraints relating them to $\hat u$. Still, we have the advantage that now all inequalities are simple bounds on individual entries of the $\xi$ vector. Moreover, we highlight that $H$ is not positive definite even though after elimination of the variables $\hat p,\hat y$ the problem is convex in $\hat u$.

Note that, by letting $f(\xi)=\frac{1}{2} \xi^\top H \xi + f^\top \xi$ be the objective function and $h(\xi)=\mathcal{C} \xi + d$, Problem~\eqref{eq:mpc_eq_and_bnd} perfect matches the setup analysed in Sec.~\ref{sec:projflcmo}. 

\subsection{Proposed Real-Time Controller}
\begin{figure*}
    \centering
    \includegraphics[width=0.65\linewidth]{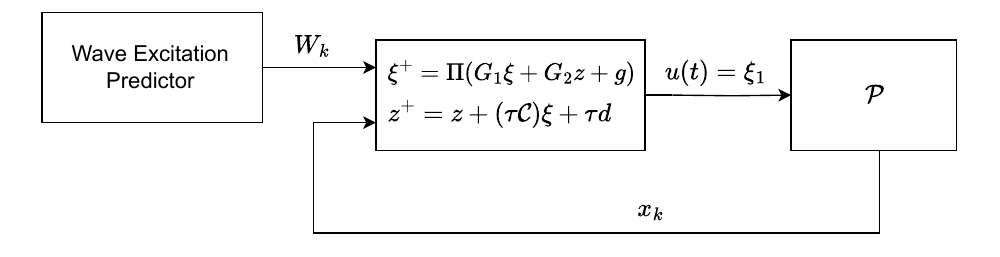}
    \caption{Feedback Control System with the proposed Real-Time Proj-FL-CMO Single-Iteration MPC}
    \label{fig:fcs}
\end{figure*}

We propose the feedback control strategy depicted in Figure~\ref{fig:fcs}. The whole state vector $x(t)$ of the WEC is assumed to be estimated with high accuracy and sampled at time instants $t = kT$, for $k\geq0$, providing samples $x_k$. The sample $x_k$ is stored in the digital device that computes the corresponding next MPC iteration over the current period. The controller is defined according to the Proj-FL-CMO dynamics \eqref{eq:projFLCMO}, i.e., the control input $u(t)$ to the WEC plant is defined, over the $k$-th sampling period, according to Algorithm \ref{alg:control}.
\begin{algorithm}
\caption{Controller actions at step $k$.}
\label{alg:control}
\begin{algorithmic}[1]
    \State Apply control input $u \gets \xi_{1}$
    \State Sample new state measurement $x_k$ and wave prediction $W_k$
    \State Compute $d$ using \eqref{eq:d_comput}
    \State Update $\xi$ and $z$ according to Eq.~\eqref{eq:projFLCMO}.
\end{algorithmic}
\end{algorithm}

\begin{remark}
    The proposed methodology is general and, in principle, may be applied to any MPC problem for LTI systems after a suitable definition of constrained outputs recasting the problem to form \eqref{eq:mpc_eq_and_bnd}. 
\end{remark}

Algorithm~\ref{alg:control} is an instance of single-iteration MPC, an emerging paradigm according to which the MPC controller is approximated by a single iteration of an algorithm solving the underlying optimization problem. The theoretical properties of single-iteration MPC schemes have been theoretically analysed in \cite[Sec. IV]{emiliano25}, in a simplified scenario, i.e., without inequality constraints and using gradient descent as the underlying algorithm. In this work, the authors establish that the stability of the control method is ensured under the condition that the sampling rate is sufficiently fast.
\begin{remark}
    A detailed analysis of the theoretical properties (i.e., stability and safety) of the proposed feedback control strategy is beyond the scope of this paper and the subject of future extended work. Yet, we conjecture that a result resembling that in \cite[Sec. IV]{emiliano25} holds true for the considered scenario.
\end{remark}

\subsection{Computational Complexity and Sample Rate Selection}
This section analyzes the computational complexity of Algorithm~\ref{alg:control} and establishes an estimate of the minimum achievable sampling rate. First, we notice that \eqref{eq:opt_probl} is equivalently rewritten as
\begin{subequations}\label{eq:projFLCMO_compact}
    \begin{align}
        \xi^+ &= \Pi( G_1 \xi + G_2 z + g + G_3 \tilde d),\\
        z^+ &= z + G_4 \xi + \tilde d,
    \end{align}
\end{subequations}
where
\begin{subequations}
\begin{align}
    G_1 &\doteq I-\tau H+\tau \C^\top (\C \C^\top)^{-1}\C \left( H+ k_p I\right) \\
    G_2 &\doteq \tau \C^\top (\C \C^\top)^{-1} k_i, \qquad G_3 \doteq  \C^\top (\C \C^\top)^{-1} k_p \\ G_4 &\doteq \tau \C, \qquad g \doteq -\tau f + \tau \C^\top (\C \C^\top)^{-1} \C f
\end{align}
and 
\begin{equation}\label{eq:d_tilde}
    \tilde d = \tau d = \mathcal{D} \begin{bmatrix}
        x_t \\ W_t
    \end{bmatrix}
\end{equation}
where, using \eqref{eq:d_comput},
\begin{equation}
    \mathcal{D} \doteq \tau \begin{bmatrix}
        C_{x,p} & C_{u,p} \\ C_{x,v} & C_{u,v}
    \end{bmatrix}.
\end{equation}\end{subequations}
Given the problem data, the data $G_1,\dots, G_4,g, \mathcal{D}$ are easily computed offline during the design phase, and thus do not impact the real-time algorithm implementation. Next, given the current controller states $\xi,z$ and measured samples $x_k, W_k$, the number of floating-point multiplications (additions and comparisons for projection are negligible in comparison) required to compute \eqref{eq:projFLCMO_compact} is summarized in Table~\ref{tab:flops_rt}. Notice that the cost of the product $G_4 \xi$ does not match the product between the dimensions of $G_4$ (i.e., $6N^2$) but is reduced to $2N^2$ thanks to the peculiar structure of $\C$, which involves identity and zero blocks.
\begin{table}[]
    \centering
    \begin{tabular}{c|c}
        Operation & FLOPs \\
        \hline
        $\tilde d$ using \eqref{eq:d_tilde} & $2N(n+N) \approx 2N^2$ \\
        \hline
        $G_1 \xi$ & $9N^2$\\
        $G_2 z$ & $6N^2$ \\
        $G_3 \tilde d$ & $6N^2$\\
        $G_4 \xi$ & $2N^2$\\
         \hline
        Total & $25 N^2$
    \end{tabular}
    \caption{FLOPs count for Algorithm~\ref{alg:control}.}
    \label{tab:flops_rt}
\end{table}

This allows for a straightforward estimation of the time required to perform a single operation. Indeed,  the minimum achievable sampling period $T^\star$ corresponds to the time required to compute one iteration of \eqref{eq:projFLCMO}, which is easily estimated as
\begin{equation}
T\geq \frac{c_{\pi}}{\textrm{OP}},
\end{equation}
where $c_\pi$ is the total cost of Algorithm~\ref{alg:control}, according to Table~\ref{tab:flops_rt}. Then, given $N = T_p/T$, we obtain the minimum rate as:
\begin{equation}\label{eq:estim_T_rt}
    T^\star = \sqrt[3]{ \frac{25 T_p^2}{\textrm{OP}} }.
\end{equation}

\section{Simulation Results}
\label{sec:simulation}

\begin{figure}[htb!]
    \centering
    \includegraphics[width=\columnwidth]{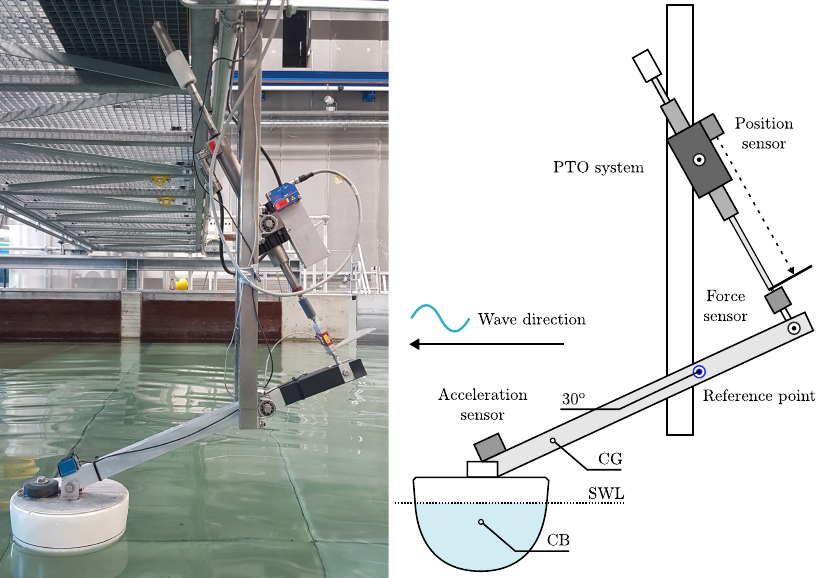}
    \caption{Photo (left) and schematic (right) of the 1:20 scale Wavestar prototype considered for performance evaluation.}
    \label{fig:wavestar}
\end{figure}

This section presents a numerical appraisal of the performance of the proposed single-iteration MPC scheme, applied to a benchmark WEC system. In particular, we consider the so-called Wavestar WEC, recently considered as a benchmark case for control evaluation within a dedicated control competition (WEC${}^{3}$OMP \cite{ringwood2023empowering}), and extensively analysed experimentally in \cite{faedo2023swell}. In particular, a 1:20 scale prototype is considered, as illustrated within Figure \ref{fig:wavestar}, as considered within the experimental setup defined in \cite{faedo2023swell}. This device features a semi-spherical floating buoy connected with a generator at the upper joint of the support arm. The dynamical model for this system, considered for simulation within this paper, is adopted from \cite{pasta2024data}, available open-source at \cite{OCEAN}. This model, which describes the behaviour of the prototype about the reference point indicated in Figure \ref{fig:wavestar}, has been computed following standard frequency-domain black-box system identification procedures\footnote{The reader is referred to \cite{pasta2024data} for further detail on the computation of the model, including \emph{e.g.} signal specification for non-parametric identification.}. 

Regarding wave generation for performance assessment, we consider an irregular sea state characterized stochastically in terms of a corresponding spectral (power) density function, described using a standard JONSWAP model \cite{hasselmann1980directional}, consistent with \cite{faedo2023swell}. Such an irregular wave condition is fully defined in terms of the significant wave height $H_w$, typical period $T_w$ and peak enhancement parameter $\gamma$. In particular, we set $H_w =$~\SI{0.0625}{\meter}, $T_w =$~\SI{1.412}{\second} and $\gamma = 3.3$, consistent with (scaled) conditions typical of the North Sea, in line with the planned location for the Wavestar system. Note that this sea state coincides with one of the operation conditions tested within the WEC${}^{3}$OMP (SS5, see \cite{ringwood2023wave}). For simulation and performance assessment, and aiming to decouple the wave estimation/forecast problem\footnote{The interested reader is referred to \emph{e.g.} \cite{pena2019critical} for further detail on estimation/forecasting of wave induced forces.} from the core of this paper (\emph{i.e.} control), we assume perfect availability of $w$ at each receding window.

Regarding computational resources, we assume that the hardware available is characterized by an OP of 100 GFLOP/s. This is a reliable estimate for average-level machines often used for real-time control (\emph{e.g.}, Speedgoat Real-time Baseline Machine, featuring a quad-core i5/i7 CPU), being consistent with the experimental setup in \cite{ringwood2023wave,faedo2023swell}. We select wave prediction of $T_p =$~\SI{2}{\second} (in accordance with the condition $T_p < T_w$), and assume that $n_i = 10$ iterations are always sufficient for IPM convergence in standard MPC. These data, using \eqref{eq:estim_T_ipm}, yields $T \approx \SI{17}{\milli\second}$. Note that this estimate is optimistic compared to the sampling rate (\SI{50}{\milli\second})  reported in the WEC${}^{3}$OMP with the very same experimental setup. This suggests that the $n_i=10$ assumption taken here is favouring the standard method. With the same device, the sampling frequency achievable by the proposed real-time solution is, according to \eqref{eq:estim_T_rt}, $T = \SI{1}{\milli\second}$.


\begin{figure}[htb!]
    \centering
    \includegraphics[width=\linewidth]{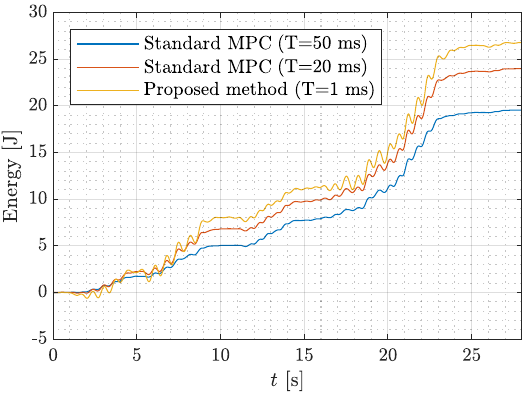}
    \caption{Energy generation performance for both the standard and the proposed single-iteration MPC.}
    \label{fig:energy}
\end{figure}

\begin{figure}[htb!]
    \centering
    \includegraphics[width=\linewidth]{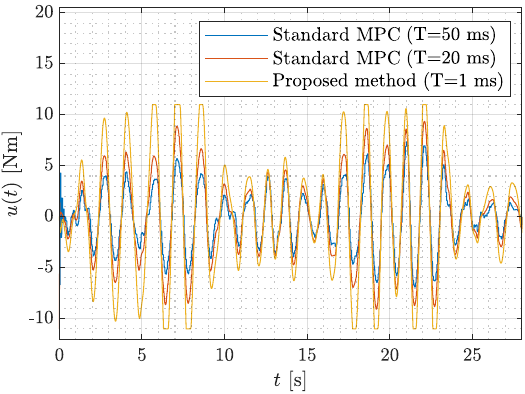}
    \caption{Control input for both the standard and the proposed single-iteration MPC.}
    \label{fig:input}
\end{figure}

Figure~\ref{fig:energy} compares the achieved performances (in terms of energy absorption) of standard MPC (ideally computing the control input of the current step instantly) running at \SI{50}{\milli\second} (standard choice) and at \SI{20}{\milli\second} (approximately the theoretical value according to \eqref{eq:estim_T_ipm}). We observe that the MPC characterized by a sampling rate of \SI{20}{\milli\second} performs considerably better than that at \SI{50}{\milli\second}, with a $22.7\%$ improvement, which provides experimental evidence for the discussion in Remark~\ref{rmk:need_small_T}. Faster rates for the standard MPC are infeasible within the given constrained hardware resources. On the other hand, the proposed solution outperforms the standard MPC: over the considered simulation horizon, $11.6\%$ more energy is produced compared to the \SI{20}{\milli\second} MPC, using the very same computational resources. 

Finally, Figure~\ref{fig:input} shows the torque applied via the power take-off system, \emph{i.e.} the applied control torque. We observe that, as expected, the saturation limits of \SI{11}{\newton\meter} are never exceeded for all controllers. Still, the control input computed with the proposed single-iteration MPC scheme better exploits the torque limit available, being compatible with theoretical optimal control results previously reported for analogous scenarios, \emph{e.g.} \cite{faedo2024experimental, merigaud2020ex}.

\section{Conclusions}
\label{sec:conclusions}

This paper addresses the energy-maximizing control problem for WECs, using a novel single-iteration MPC approach. In particular, we propose an efficient, first-order algorithm (Proj-FL-CMO) by building upon recent literature on control-based optimization algorithm design, and study its theoretical convergence properties. 

Next, we recast the WEC MPC problem to equality-constrained optimization with bounds and apply Proj-FL-CMO to define the single-iteration MPC controller. The simulation results obtained on a benchmark Wavestar WEC in typical sea conditions demonstrate the ability to deliver superior performances when compared to standard MPC, while preserving real-time capability within limited computational resources.

\balance
\bibliographystyle{ieeetran}
\bibliography{references}

\end{document}